\documentclass{article}
\usepackage{graphicx}

\usepackage[usenames]{color}
\usepackage{makeidx}

\begin{document}

\title{{\bf Macroscopic entanglement between wave-packets at finite temperature}}

\author{Ana M. Martins\\
Departamento de F\'{i}sica, Instituto Superior T\'{e}cnico, 1049-001
Lisboa, Portugal}

\date{\today}

\maketitle

\begin{abstract}

We investigate entanglement between collective operators of two wave-packets of finite spectral bandwidth, in two different approximations of the Multimode Parametric-Down Conversion  (MPDC) process: the {\it pairwise } and the {\it one-to-all} interaction patterns. For collective operators we choose the macroscopic amplitudes of each wave-packet defined by the Fourier Transform of their microscopic mode amplitudes. This approach intends, to respond to realistic experimental conditions, where measurements apparatuses may not resolve single microscopic mode amplitudes but rather the collective amplitude of the wave-packets. To quantify the bipartite macroscopic entanglement we use the logarithmic negativity. We relate the time dependent degree of macroscopic entanglement with the complexity (number of modes and interaction pattern) and the temperature of the system. Our results show that  the macroscopic entanglement increases linearly with the number of micro-modes in the case of the {\it one-to-all} interaction, while in the {\it  pairwise} interaction it is constant. Moreover, in the {\it one-to-all} pattern the {\it birth time of entanglement} and the critical temperature decrease with increasing the number of micro-modes. We draw the graphs associated with the two interaction patterns and related the degree of collective entanglement with the connectivity and the index of each vertex (mode) of the graph. We conclude that quantum information and computation tasks may be achieved more efficiently by manipulating appropriated collective operators in some macroscopic systems,  then by using their microscopic counterparts.

\end{abstract}

PACS number(s)  03.67.Bg, 42.65.Yj, 03.67.-a


\newpage

\section{Introduction}

In the last years a great number of experimental and theoretical research has been dedicated to find out how to create and extract entanglement from macroscopic interacting systems \cite{Vedral2008}. Between them are chains of interacting spins \cite{Vedral2001,Briegel2005}, harmonic oscillators and harmonic lattices \cite{Vedral2006, Anders2008,Vedral2010} and macroscopic states of light \cite{Martini2010, Leuchs2010}. These works intend, to respond to several aspects of the physical reality. 

In one hand, in some realistic experimental conditions the measurements apparatuses may not resolve single constituents composing the systems but rather global properties, thermodynamically speaking, they measure macroscopic observables rather than microscopic ones. 
In this case, it is quite pertinent to ask wether the quantum correlations existent between the microscopic constituents of the systems will survive when the number of these constituents increases? On the other hand, thermal states of interacting macroscopic systems, at finite temperature, are real classical states and the following question arises naturally: can interacting systems, initially in a thermal state, evolve to quantum entangled states? 

These kind of issues relaunch the old question of the passage from quantum to classical behavior with increasing complexity of the systems and with increasing temperature \cite{Kofler2008}.

Beside this pure fundamental issue, macroscopic entanglement may become an important physical resource to perform communication and computation tasks with efficiency not achievable classically \cite{Chuang},  with the great advantage (comparing with microscopic entanglement) of being created and manipulated in simple and accessible macroscopic systems, at room temperature.

Parametric down-conversion (PDC) processes have been used to visualize the quantum to classical transition at the single-mode single-photon level   \cite{Walls1994}. Experimental techniques, developed recently \cite{Leuchs2010, Leuchs2012}, detected quantum entanglement between bright squeezed vacuum beams (containing about $10^5$ photons). This suggests that multimode PDC interactions are good candidates to study the joint effect of complexity and finite temperature, in macroscopic entanglement. 

Our perspective in this work will be to identify a small number of collective observables (true physical quantities) \cite{Vedral2006}, that depend in the size (number of microscopic constituents) of the macroscopic systems and such that their measurement give information about the degree of macroscopic entanglement. We choose for collective observables the macroscopic amplitudes of two wave-packets of finite spectral bandwidth, generated in a multimode parametric down conversion process, in two different approaches: the {\it pairwise} and the {\it one-to-all} interactions. 

These macroscopic amplitudes are extensive quantities scaling with the number of the microscopic modes of the wave-packets, and are defined by the Fourier Transform of the amplitudes of the microscopic modes. The corresponding observables are macro Bose operators of a two dimensional continuous variable Hilbert space.

To measure the time dependent degree of entanglement of the bipartite macroscopic wave-packets we use the {\it logarithm negativity} and analyze its behavior in terms of the size of the wave-packets, and of the temperature of the nonlinear medium where the interaction takes place. 

To give a pictorial representation of the physical interactions, we associate a graph  \cite{Briegel2006} with each interaction pattern and show that they are fundamentally different: while the graph of the {\it pairwise} interaction is composed of $n$ disconnected subgraphs whose vertices have degree $1$, the graph of the {\it one-to-all} interaction is completely connected and its vertices have degree equal to the number $n$ of micro-modes in each wave-packet. This structural difference between the two graphs will be reflected in the behavior of the degree of macroscopic entanglement. We show that more complex is the system stronger is the macroscopic entanglement and higher is the critical temperature where it disappears. Another interesting conclusion is that, at finite temperatures, the minimum interaction time needed for the wave-packets to be entangled decreases with increasing their complexity.

Our work extends previous studies with bright beams in three aspects: (a) our concept of a macroscopic system resides in the number of  interacting microscopic modes (degrees of freedom), rather than on the amount of photons produced during the interaction, and is valid for any finite number of micro-modes in each-wave packet. (b) entanglement between macroscopic systems is detected through collective operators that retain the global properties of the bipartite Gaussian, continuous variables macro-modes rather than computing multimode entanglement, (c) the effect of the temperature of the nonlinear medium on entanglement is taken into account, previous results only consider the initial vacuum state.

The plan of the paper is the following: In Sec.2 we introduce the Hamiltonean of the nonlinear interacting wave-packets in two different approaches and derive their dynamical behavior. In Section 3 we define the time-dependent collective amplitudes in terms of the microscopic ones. In Section 4 we compute the macroscopic degree of entanglement and the birth time of entanglement in terms of the number of micro-modes and of the temperature. Finally in Section 5 we present our conclusions.

\section{System dynamics }

In the broadband down-conversion process \cite{Bloembergen1965} the pump wave {\bf 0}, the signal {\bf 1} and the idler {\bf 2}, behave like short wave-packets of bandwidths $\Delta \omega_0$, $ \Delta \omega_1$ and $\Delta \omega_2$, respectively, interacting in a nonlinear medium.  

In a discrete mode description of the electromagnetic field (e.m.) \cite{Yariv1967} each wave-packet ${\bf j}$, is composed by $n=2m+1$ discrete monochromatic modes of frequency $\omega_{j,k} = {\bar \omega}_j + k \delta$ ( ${\bf j}=0,1,2 \,\, ;k=-m,-m+1,...,m-1,m$) where $ {\bar \omega}_j$ is the central frequency of wave-packet ${\bf j }$ and $\delta = \frac{\Delta \omega_j }{2m}$ is the frequency spacing between two neighbor modes.

We assume that the energy conservation condition is obeyed by the central frequencies of each wave-packet, i.e., $ {\bar \omega}_0= {\bar \omega}_1 + {\bar \omega}_2$ and we also assume that the spectral bandwidth of wave ${\bf j}$, $\Delta \omega_j = \omega_{j,m} - \omega_{j,-m} = 2 m \delta$, is small enough to prevent overlapping of the three waves and such that each mode interacts nonlinearly with all the others, except for those belonging to the same wave-packet. The quantum version of the classical Hamiltonean \cite{Weiland1977, Martins1988} describing this nonlinear interaction is 
\begin{equation}\label{3waves}
{\hat H }_{NL} = \sum_{ j=0}^{2} \sum_{k=-m}^{m} \hbar \omega_{j,k} {\hat a}_{j,k}^{\dag} {\hat a}_{j,k} + i  \hbar g \sum_{p=-m}^{m} \sum_{k=-m}^{m} \sum_{l=-m}^{m} ({\hat a}_{0,p}{\hat a}_{1,k}^{\dag}{\hat a}_{2,l}^{\dag} -h.c.)
\end{equation}
where the coupling constant $g $ is proportional to the second order susceptibility of the medium. 

When the incident pump wave, is intense, the ensemble of modes $({\bf 0},k)$ composing the pump wave-packet can be treatead classically as a coherent undepleted field of complex amplitude  $\alpha_0 = | \alpha_0| e^{-i 2 \varphi} $ and an arbitrary pump phase $\varphi $. Within this approximation we obtain the following {\it one-to-all }Multimode Parametric Down Conversion (MPDC) Hamiltonean, 
\begin{equation}\label{2waves}
{\hat H }_{1} = \sum_{ j=1}^{2} \sum_{k=-m}^{m} \hbar \omega_{j,k} {\hat a}_{j,k}^{\dag} {\hat a}_{j,k} + i w \hbar \sum_{k=-m}^{m} \sum_{l=-m}^{m} ( e^{- i({\bar  \omega}_0 t + \varphi )}{\hat a}_{1,k}^{\dag}{\hat a}_{2,l}^{\dag} -h.c.) 
\end{equation}
where each mode $({\bf 1},k )$ of the signal wave interacts with every mode $({\bf 2}, l )$ of the idler wave and vice-versa. In this approximation the coupling parameter $w=g |\alpha_0|$ is proportional to the amplitude of the pump and $t$ is the interaction time that is proportional to the length of the propagation path of the e.m. field in the nonlinear medium.

When $\Delta \omega_0 <<  \Delta \omega_1 , \,\  \Delta \omega_2 $, MPDC Hamiltonean (\ref{2waves}), can be further simplified,  since the coupling between the energy non conserving modes can be neglected and each mode $({\bf 1},k)$ of the signal interacts uniquely with mode $({\bf 2},-k)$ of the idler, such that $\omega_{1,k} + \omega_{2,-k} =  {\bar \omega}_0$. Within this new approximation the Hamiltonian  ${\hat H }_{1}$ reduces to the {\it pairwise} MPDC Hamiltonian \cite{Braunstein1988(1), Paris2009}
\begin{equation}\label{2waves1}
{\hat H }_{2} =  \sum_{k=-m}^{m} {\hat H_k }  
\end{equation}
and
\begin{equation}\label{PDC}
{\hat H_k } =  \hbar \omega_{1,k} {\hat a}_{1,k}^{\dag} {\hat a}_{1,k}  +\hbar \omega_{2,-k} {\hat a}_{2,-k}^{\dag} {\hat a}_{2,-k} + i w \hbar ( e^{- i({\bar  \omega}_0 t + \varphi )}{\hat a}_{1,k}^{\dag} \,\ {\hat a}_{2,-k}^{\dag} -h.c.) 
\end{equation}
is the Hamiltonian of each pair of field modes interacting under the PDC approximation \cite{Louisell1961}.

A better insight about the type of interaction between the microscopic modes of the two wave-packets can be achieved when a graph is associated to the interaction Hamiltonians ${\hat H }_{1} $ and ${\hat H }_{2} $, as we do in Fig.1. In panel (a) graph $G_1=(V,E_1)$ represents the interaction hamiltonean ${\hat H }_1$ and in panel (b) graph $G_2=(V,E_2)$ is associated with Hamiltonian ${\hat H }_2$. In both graphs each vertex $ ({\bf j},k) \in V= \{({\bf  j},k) :{\bf j }={\bf 1},{\bf 2} \mbox{ and } k=-m,-m+1,...,m-1,m \}$ represents a microscopic mode of frequency $\omega_{j,k}$ and each edge $\{ ({\bf 1},k),({\bf 2},l) \} \in  E$ represents the interaction between mode $({\bf 1},k)$  of wave {\bf 1} with mode $({\bf 2},l)$ of wave {\bf 2}. The coupling strength $w$ is the same for all edges.

 The degree of vertex $({\bf j},k)$ is the number ${\cal N}_{j,k}$ of its neighbors \cite{Briegel2006}. In graph $G_1$, ${\cal N}_{j,k} =n$, increases with the number $n$ of modes in each wave-packet and, even though micro-modes of the same wave-packet do not interact directly, a common micro-mode in the other wave-packet is an interaction mediator between them, therefore, there is at least a path connecting any pair of vertices $\{ ({\bf j},k) , ({\bf i},l)   \}, $ in graph $G_1$, i.e.,  $G_1$ is a connected graph. In graph $G_2$, ${\cal N}_{j,k} =1$ for any vertex and $G_2$ is composed by $n$ similar, disconnected subgraphs. These differences between the degree of the vertices and the connectivity of the two graphs must be reflected in the behavior of the degree of entanglement of the two wave-packets as we will show.

\begin{figure}[htb]
\centering
\includegraphics{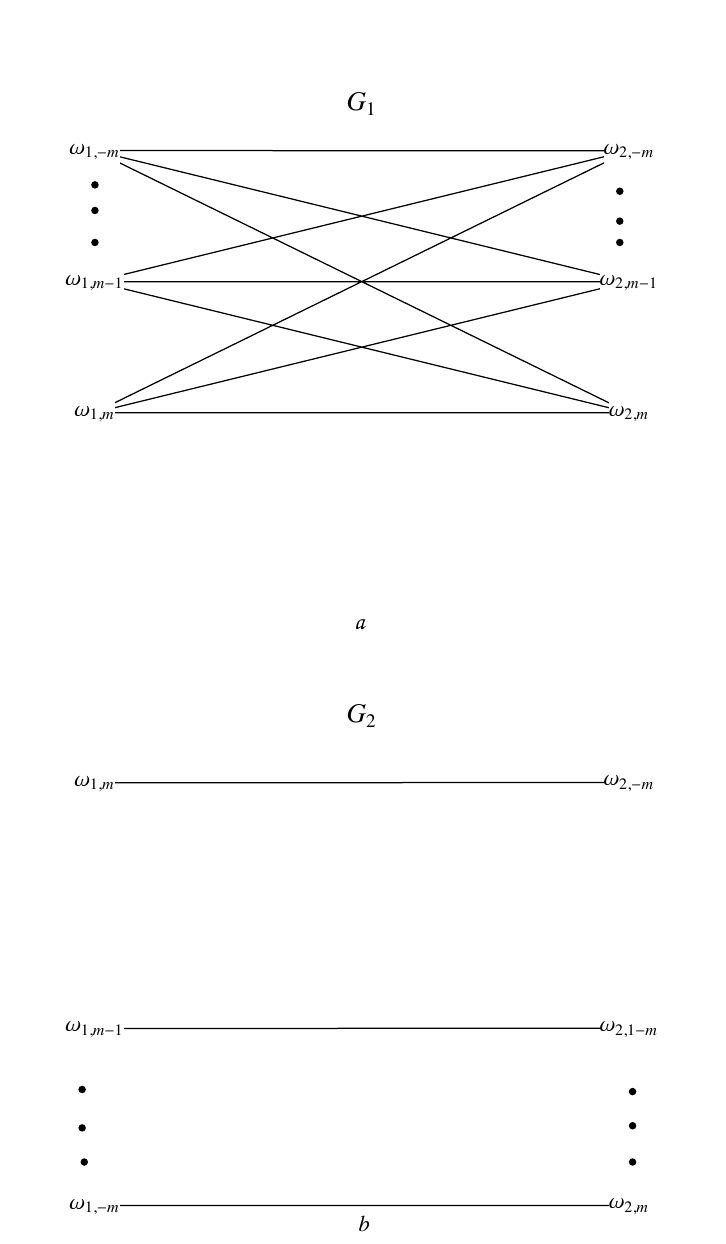}
\caption{\it 
Graphs $G_1$ and $G_2$ represent the interaction between the microscopic modes of wave-packets  {\bf 1} and {\bf 2} for Hamiltonians ${\hat H }_1$, panel ({\bf a}), and ${\hat H }_2$, panel {\bf (b)}.  Each vertex represents a microscopic mode of frequency $\omega_{i,j}$ and each edge represents the interaction between mode $({\bf 1},k)$  of wave {\bf 1} with mode $({\bf 2},l)$ of wave {\bf 2}. The coupling strength $w$ is the same for all edges. }
\label{FIG.1}
\end{figure}

\section{Collective Operators}

When measuring wave-packets composed by several micro-modes, the measuring apparatus may not resolve and detect the individual mode amplitudes but rather the macroscopic amplitude of each wave-packet. In this case, choosing the collective operators as the physical quantities actually detected by the measuring apparatus, will enable us to relate the degree of macroscopic entanglement between wave-packets with the number of their microscopic constituents.

The frequency dependent collective operators associated with the macroscopic amplitudes of wave-packets {\bf 1} and {\bf 2}, at the initial time $t=0$, are naturally defined by the Fourier Transform of the $(n=2m+1)$ modes of each wave-packet
\begin{equation}\label{macro}
{ \hat A}_{{\bf 1}0} = \frac{1}{\sqrt  {2m+1}} \sum_{k=-m}^{m}  {\hat a}_{{\bf 1}0,k} \,\ ; \,\,\  { \hat A}_{{\bf 2}0}^{\dag} = \frac{1}{\sqrt  {2m+1}} \sum_{k=-m}^{m}  {\hat a}_{{\bf 2}0,k}^{\dag}
\end{equation}
the subscript zero refers to initial time. These collective amplitudes obey the commutation relations $[ { \hat A}_{{\bf i}0}, { \hat A}^{\dag}_{{\bf j}0}] = \delta_{ij} $ and $i,j=1,2$.

The time dependent collective amplitude operators are simply given by the dynamical evolution of the initial amplitudes defined above, i.e., $ { \hat A}_{{\bf j}}(t) = \hat{U}_t^{\dag}{ \hat A}_{{\bf j}0} \hat{U}_t  \,\ j=1,2 $ where  $\hat{U}_t $ is the unitary time propagator of the composite system, that preserves the canonical commutation relations at any time $t$. 

Let us stress that the collective amplitude operators $ { \hat A}_{\bf j }(t)$ are true annhilation operators of the {\it macro-modes} (wave-packets) {\bf 1} and {\bf 2}. The macroscopic quantum number operators associated with these macro-modes are naturally given by ${ \hat N}_{\bf j}={ \hat A}_{\bf j}^{\dag} { \hat A}_{\bf j }$ and their  eigenvectors are collective Fock states $| N_{\bf j} \rangle , \,\ j=1,2$ . 

Let 
\begin{equation}\label{Macroquadratures}
\hat { Q}_{\bf j} = \frac{1}{{\sqrt  {2 }}}({\hat {A}_{\bf j} }  + {\hat {A}_{\bf j}^{\dag}  } ) \,\,\, ; \,\,\,\  \hat {P}_{\bf j}  =  - \frac{i}{{\sqrt  {2 }}}({\hat { A}_{\bf j} }  - {\hat {A}_{\bf j}^{\dag}  } )
\end{equation}
be the related quadratures phase collective operators of {\it macro-mode} {\bf j}.

Grouping  together the operators $\hat { Q}_{\bf j}$ and ${ \hat  P}_{\bf j}$ in a vector of operators $  {\hat X}  \equiv  ( {\hat Q}_{\bf 1} , {\hat P}_{\bf 1} , {\hat  Q}_{\bf 2},  {\hat P}_{\bf 2}  )^T$, then the canonical commutation relations for the $ {\hat X}_j^{'}$ s at any time $t$ are encoded in the symplectic form $[ {\hat X}_i,  {\hat X}_j]= i \Omega_{i,j}$, where $\Omega_{i,j}$  are the elements of the symplectic matrix
\begin{equation}\label{simplectic}
{\bf  \Omega}= \bigoplus_{k=1}^{2} \left( 
\begin{array}{cc}
0  &1 \\ 
-1   & 0 \\ 
\end{array}
\right) 
\end{equation}
The initial and time-dependent quadratures phase collective operators are then related by the symplectic structure of the dynamical transformation.

The time dependent  amplitudes of the micro-modes, in the {\it pairwise} interaction, are given by eqs. (\ref{PDC3}) and (\ref{PDC4}) of Appendix A. Then, the corresponding time dependent collective amplitudes have the simple form
\begin{equation}\label{MacrotimeA}
 { \hat A}_{\bf 1}(t) =   \frac{\cosh wt  }{\sqrt  {2m+1}} \sum_{k=-m}^{m}e^{-i \omega_{1,k} t} \,\ \hat{a}_{{\bf 1}0,k}  +\frac{\sinh wt  }{\sqrt  {2m+1}}\,\ \sum_{k=-m}^{m}e^{{-i \omega_{1,-k}  t}} \,\ \hat{a}_{{\bf 2}0,-k}^{\dag}
\end{equation}  
\begin{equation}\label{MacrotimeB}
  { \hat A}_{\bf 2}^{\dag}(t) =  \frac{\sinh wt  }{\sqrt  {2m+1}} \sum_{k=-m}^{m} e^{i \omega_{2,-k}  t} \,\ \hat{a}_{{\bf 1}0,k} + \frac{\cosh wt  }{\sqrt  {2m+1}}\,\ \sum_{k=-m}^{m} e^{i \omega_{2,-k}  t} \,\  \hat{a}_{{\bf 2}0,-k}^{\dag}
\end{equation}
In the case of the {\it one-to-all} interaction, the time dependent collective amplitudes are
\begin{equation}\label{MacrotimeA1}
  { \hat A}(t) \equiv  { \hat A}_{\bf 1}(t) =  \sum_{j=1}^{2m+1} m_j (t) \hat{a}_{j0} + \sum_{j=2m+2}^{4m+2} n_j (t) \hat{b}_{j0}^{\dag} 
\end{equation} 
\begin{equation}\label{MacrotimeB1}
{ \hat B}^{\dag}(t) \equiv  { \hat A}_{\bf 2}^{\dag}(t) = \sum_{j=1}^{2m+1} t_j(t) \hat{a}_{j0} + \sum_{j=2m+2}^{4m+2} u_j (t) \hat{b}_{j0}^{\dagger}
  \end{equation} 
where we used the time dependent amplitudes of the micro-modes, eqs.(\ref{motion00}), (\ref{motion01}) derived in the Appendix A. Note that the Bose operators $\hat{a}_{{\bf j}0,k}$ were renamed for commodity of calculations. The expressions of the coefficients $m_j(t)$, $n_j(t)$, $t_j(t)$ and $u_j(t)$, are given by equations (\ref{Coefficient1}) and (\ref{Coefficient2}) of Appendix B.

\section{Degree of macroscopic entanglement}

In this work, we are interested in the amount of entanglement we can extract from the system if only the averaging collective observables $ {\hat Q}_{\bf 1,2} $ and $ {\hat P}_{\bf 1,2} $ were measured and manipulated. In particular, we compute the degree of macroscopic entanglement in terms of: (a) the number $n$ of modes constituting each wave-packet, (b) the interaction time $t$, (c) the strength of the coupling parameter $w$ and (d) the equilibrium temperature $T$.

Two physical situations, described by two different initial quantum states, have special interest when considering macroscopic correlations and entanglement in multimode parametric down conversion processes. They are the spontaneous emission and the pair production at temperature $T$.  

In the MPDC spontaneous emission each mode $({\bf j},k)$ composing the two wave-packets is initially in the vacuum state $ |0 \rangle_{{\bf j}0,k}$ and the initial state of the composite system is the pure and separable state given by the tensor product $ | \psi \rangle_0 = \bigotimes_{{\bf j}=1}^{2}  \bigotimes_{k=-m}^{m} |0 \rangle_{{\bf j}0,k}$.

When the nonlinear medium where the waves propagate is in thermal equilibrium at temperature T, the initial state of the wave-packets is a separable mixed state given by the density operator ${\hat \rho}_0 = \bigotimes_{{\bf j}=1}^{2}  \bigotimes_{k=-m}^{m} {\hat \rho}_{{\bf j}0,k}$
where $ {\hat{ \rho}}_{{\bf j}0,k}  = \sum_{n_{{\bf j0},k} =0}^{\infty} e^{- \beta_{{\bf j},k} n_{{\bf j0},k} } | n_{{\bf j0},k}\rangle \langle n_{{\bf j0},k} |$
is the thermal field density operator of mode $({{\bf j},k})$, with $ \beta_{{\bf j},k} =\hbar \omega_{{\bf j},k} / (k_B T) $ and $k_B$ is the Boltzmann constant. The initial mean number of photons in mode $({{\bf j},k})$ is ${\bar n}_{{{\bf j0},k}}=1/(e^{\beta_{{\bf j},k}}-1)$. At room temperature and in the optical part of the electromagnetic spectrum ${\bar n}_{{{\bf j},k}}<<1$, however, in the microwave part of the spectrum ${\bar n}_{{{\bf j},k}}>>1$ and we cannot ignore the presence of thermal photons in the nonlinear medium.  Both initial states we are considering are Gaussian states.

Since the Hamiltonians ${\hat H}_1$  and ${\hat H}_2$ are bilinear in the field macro-modes, the overall output state is also Gaussian and the separability criterion \cite{Simon2000,Werner01}  is completely characterized by the first and the second statistical moments of the collective quadrature phase operators, which will be denoted, respectively, by the vector of first moments $ {\bar X } \equiv  ( \langle {\hat { Q}_{\bf 1} } \rangle ,\langle {\hat { P}_{\bf 1}  } \rangle,  \langle {\hat  Q}_{\bf 2}  \rangle  , \langle {\hat { P}_{\bf 2}  } \rangle )$ and by the covariance matrix (CM) ${\bf \Sigma } $ of elements 
\begin{equation}\label{covariance}
{\bf \Sigma}_{ij} \equiv  \frac{1}{2} \langle  {\hat { X}_i  } {\hat {X}_j } + {\hat {X}_j  } {\hat { X}_i } \rangle - \langle  {\hat { X}_i  } \rangle  \langle {\hat {X}_j } \rangle
\end{equation} 
The first moments can be adjusted to zero by local unitary operations which leave invariant entropy and entanglement.

The time dependent covariance matrix $\mathbf {\Sigma} (t)$  for the collective observables can be expressed in terms of the three $(2 \times 2)$ block matrices $ \mathbf { \alpha }(t)$, $ \mathbf { \beta }(t)$ and $ \mathbf { \gamma }(t)$
\begin{equation}\label{macrovar}
{\bf \Sigma }(t) = \left( 
\begin{array}{cc}
\alpha (t)  & \gamma (t)  \\ 
\gamma^T(t)    & \beta (t) \\ 
\end{array}
\right)
\end{equation}
The diagonal blocks  $ \mathbf { \alpha }(t)$, $ \mathbf { \beta }(t)$ are the local CM of wave-packets {\bf 1} and {\bf 2}, respectively. The off-diagonal block, $ \mathbf { \gamma }(t)$, encode the intermodal correlations (quantum and classical) between the two wave-packets.

The entries $ \Sigma_{ij} $, in the {\it one-to-all} interaction, computed in the Appendix B, (eqs.(\ref{sigma11}) - (\ref{sigma23})), depend on the number $n$ of micro-modes in each wave-packet. Making $n=1$ in these equations we obtain the entries of the macroscopic CM for the {\it pairwise} interaction, given by eqs.(\ref{sigma11p})-(\ref{sigma23p}), which are independent of the number $n$ of micro-modes. This result is not surprising, since we have seen that the graph $G_2$ associated to this interaction is simply a collection of $n$ disconnected subgraps, each of them representing the well known two-mode PDC process.

Hamiltonians that are quadratic in the bosonic operators, possess universal quantum invariants, i.e., certain combinations of variances which are conserved in time independently of the concrete form of coefficients of the Hamiltonian \cite{Illuminati04,Serafini04}. These invariants exist due to the symplectic structure of the transformation relating initial and time-dependent values of the quadrature components operators.  Concerning the statistical moments, these universal invariants are \cite{Dodonov2005}
\begin{equation}\label{invariants}
{\cal I}_1 = \det {\bf \Sigma} (t)\,\ ;  \,\,\,\,\,\,\      {\cal I}_2 = \det \alpha (t)+ \det \beta (t)+ 2 \det \gamma (t)
\end{equation} 
Defining the quantities 
\begin{equation}\label{st}
{\cal S} (t) = {\cal S}_0 + \frac{1}{2} (\det \gamma (t) - |\det \gamma (t)| )\,\ ;  \,\,\,\,\,\,\      {\cal S}_0= {\cal I}_1  -   \frac{1}{4}{\cal I}_2  +\frac{1}{16}
\end{equation} 
the entanglement criterion can be written in a simple form
\begin{equation}\label{criterion}
{\cal S} (t) < 0
\end{equation} 
For both vacuum and thermal initial states and for both interaction patterns, we prove in the Appendix B, that 
\begin{equation}\label{simetria1}
{\cal S} (t) = S_0 -  | \det \gamma (t)| 
\end{equation} 
with
\begin{equation}\label{S0thermal}
{\cal S} _0= { \bar N}_{10}{ \bar N}_{20} ( { \bar N}_{10}+1) ( { \bar N}_{20}+1)
\end{equation}
where ${ \bar N}_{10}$ and ${ \bar N}_{20}$ are the initial average number of photons in each wave-packet, (see eq.(\ref{N10p})). For the initial vacuum state, ${ \bar N}_{10}= {\bar N}_{20}=0 $ and ${\cal S}_0=0$.

\subsection{Birth time of entanglement}

Using the entanglement criterion (\ref{criterion}), we conclude that the two wave-packets are entangled for interaction times $t_i$ obeying
\begin{equation}\label{ent}
|\det \gamma (t_i)| > {\cal S} _0
\end{equation}
where $|\det \gamma (t_i)| $ is an increasing function of  $t$. The instant $t_{\cal E}$, such that $|\det \gamma (t_{\cal E})| 
= {\cal S}_0 $ is the minimum interaction time needed before macroscopic entanglement appears, for obvious reasons it will be named the {\it birth time of entanglement} (BTE). 

\begin{figure}[htb]
\centering
\includegraphics{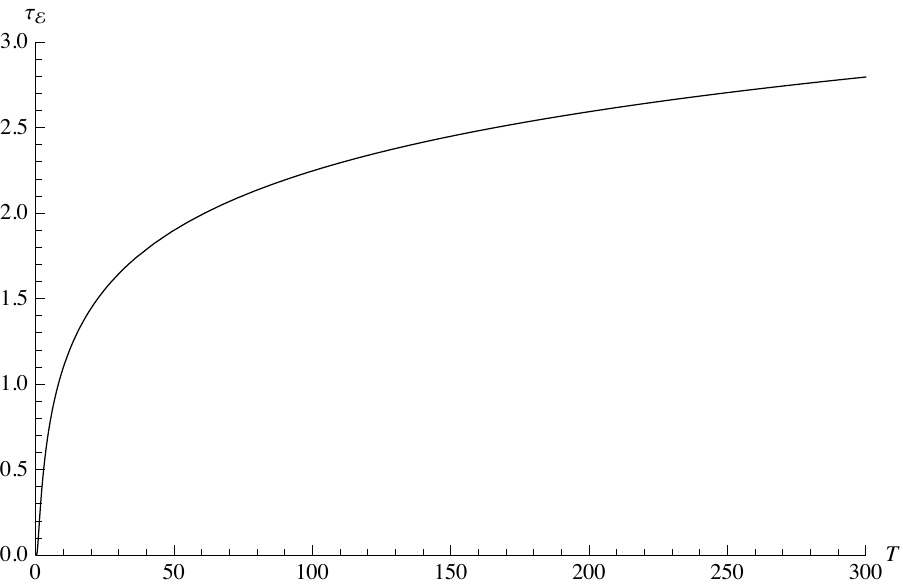}
\caption{\it 
Dimensionless BTE, $\tau_{\cal E}$, over the temperature, for the dimensionless frequencies $ { \bar \omega}_1 = \omega_1/g=200 ,\,\ {\bar \omega}_2=\omega_2/g= 400$ and for different values of the dimensionless mismatch parameter $y$: $y =0$ (triangles), $y =0.5$ (circles) and $y =0.9 $ (stars).}
\label{FIG.1}
\end{figure}

For the vacuum initial state and for both {\it pairwise} and {\it one-to-all} interactions, ${\cal S} (t) >  0$ for $t > 0$, therefore the two wave-packets are entangled since the very beginning of the interaction. On the contrary, for an initial thermal state, the condition (\ref{ent}) is attained only after a finite time of interaction, $t_i \equiv t_{\cal E}$, as passed. 

For the {\it pairwise} interaction the {\it birth time of entanglement}
\begin{equation}\label{birth}
t_{\cal E}= \frac{1}{2 w} \mbox{ ArcSinh}  \left\{ 2 \frac{{\cal S} _0^{1/2}}{{ \bar N}_{10} +{ \bar N}_{20}+1} \right\}
\end{equation}
is independent of the number $n$ of micro-modes. It depends on the temperature $T$, through the initial average  macroscopic number of photons. The dependence of de dimensionless BTE, $\tau_{\cal E}=w t_{\cal E}$, given by eq.(\ref{birth}), over the temperature is displayed in Fig.2, for the dimensionless frequencies $ { \bar \omega}_1 =200 ,\,\ {\bar \omega}_2=400$ and dimensionless spectral bandwidths $\Delta { \bar \omega}_1 = \Delta { \bar \omega}_2 =0.02$. We observe that BTE increases with the temperature although this growth slows down at higher temperatures.

\begin{figure}[htb]
\centering
\includegraphics{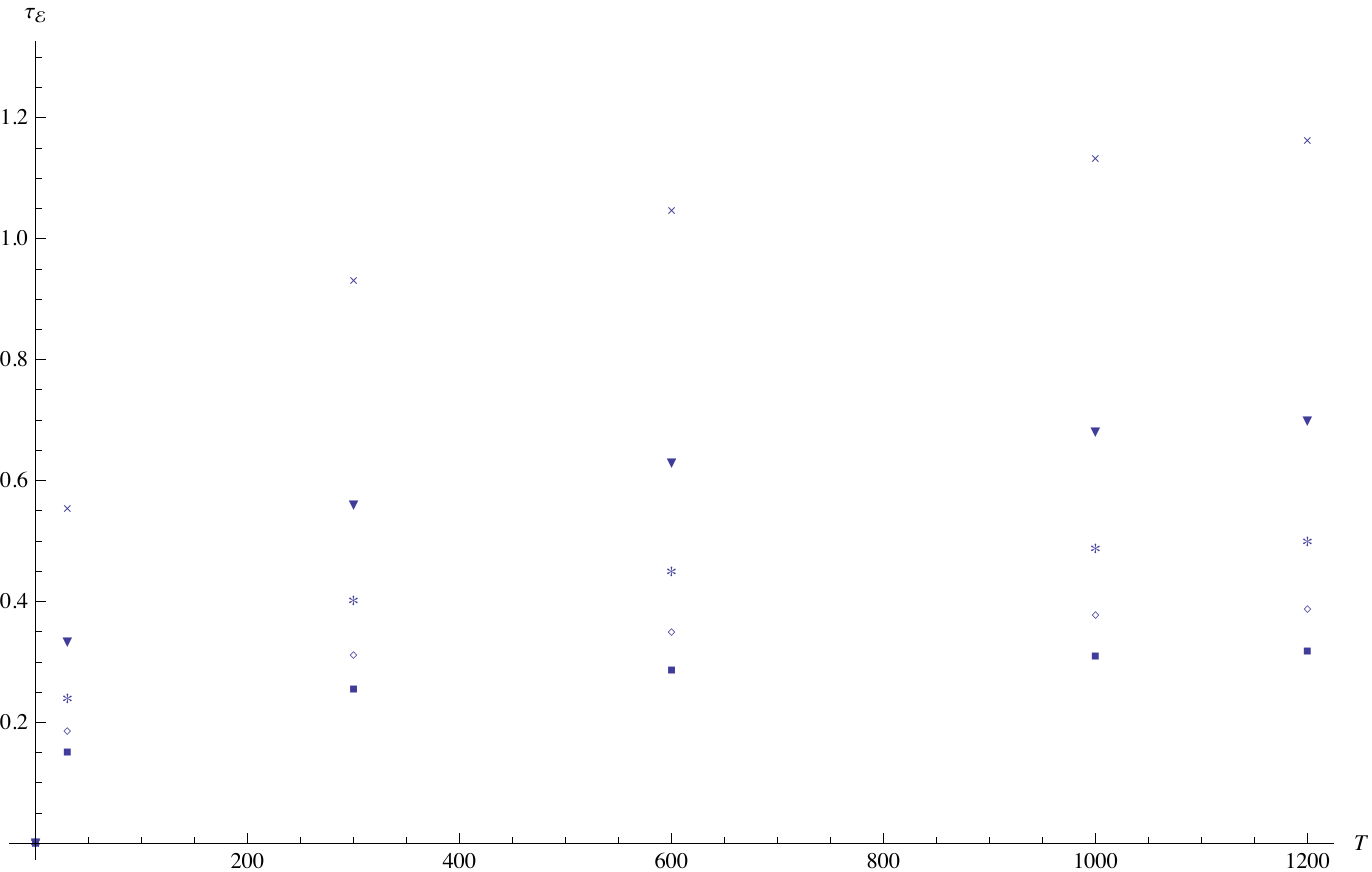}
\caption{\it 
Dimensionless BTE, $\tau_{\cal E}$ in the case of the {\it one-to-all} interaction is displayed as a function of the temperature $T$ for $n$ modes in each wave-packet: $n=3$ (crosses), $n=5$ (inverted triangles) , $n=7$ (stars), $n=9$ (diamonds) and $n=11$ (triangles). We assume the same dimensionless frequencies as in Fig.2.}
\label{FIG.3}
\end{figure}

A similar behavior is found in the case of the {\it one-to-all} interaction, see Fig.3, where the BTE, $\tau_{\cal E}$, is displayed as a function of the temperature $T$ for different number of micro-modes. We observe that for a given $n$ the BTE increases with the temperature. We can understand the behavior of the BTE with changing temperature in the following way. The MPDC interaction produces correlated pair of photons in waves {\bf 1} and {\bf 2} and entanglement between the wave-packets starts only when the system attains the {\it break-even}, this is, when the average energy of the generated pair exceeds the average energy of the initial thermal state.

The effect of the number of micro-modes in the {\it birth time of entanglement} is displayed in Fig.4 for different temperatures. We observe that, for a given temperature $T$, the  {\it birth time of entanglement} decreases with increasing number of modes in each wave-packet.

\subsection{Logarithmic negativity}

To quantify entanglement between the macroscopic wave-packets we use the {\it logarithmic negativity} ${ \cal E_N}$ \cite{Serafini04} given by
\begin{equation}\label{lnega}
{ \cal E_N} = max[0,-ln \,\ { \tilde \nu }]
\end{equation}
where ${ \tilde \nu }$ is the symplectic eigenvalue of the partially transposed Covariance Matrix ${\tilde {\bf \Sigma}}$. 

Since the CM in the {\it pairwise} interaction is independent of the number $n$ of micro-modes and since it coincides with the CM of the {\it one-to-all} interaction for $n=1$, we conclude that the macroscopic degree of entanglement ${ \cal E_N}$, in the {\it pairwise} interaction, is $n$ independent and its value is the same as if there were a single pair of modes in each wave-packet, provided that their frequencies are equal to the central frequencies of the two wave-packets and their average number of photons are equal to the average number of {\it macroscopic photons} in each wave-packet.  

On the contrary, in the {\it one-to-all} interaction, the complexity of the interaction pattern is revealed on the macroscopic degree of entanglement. We  compute ${ \cal E_N}$ as a function of the number $n$ of micro-modes in Fig.5,  for the initial vacuum state (triangles) and for a thermal state at equilibrium temperature $T= 30^{\circ} K$ (circles) and for the dimensionless interaction time $\tau_i =0.3324$. The same dimensionless frequencies as in Fig.2 are assumed. The solid lines are parallel straight lines. We conclude that for, a given temperature, the degree of entanglement increases linearly with the number of micro-modes $n$, in each wave-packet provided that the time of interaction is bigger than the BTE  corresponding to that temperature. The value of the slope of the straight lines is an increasing function of the coupling parameter $w$. 

\begin{figure}[htb]
\centering
\includegraphics{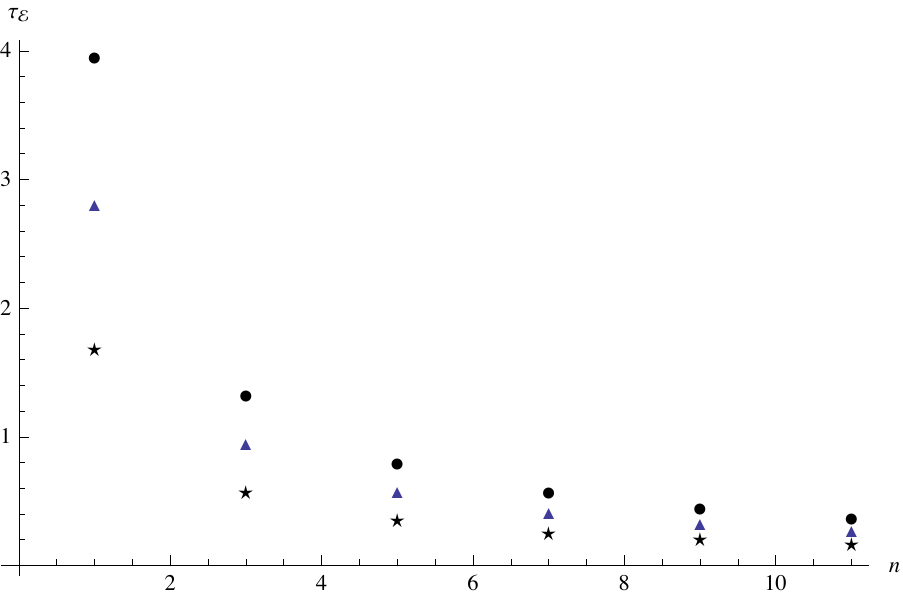}
\caption{\it 
Dimensionless {\it birth time of entanglement}, $\tau_{\cal E} $ is plotted as a function of the number $n$ of modes in each wave-packet for equilibrium temperatures: $T=30 ^{\circ} K$ (stars), $T = 300 ^{\circ} K$ (triangles), and $T= 3000 ^{\circ} K$ (circles). We assume the same dimensionless frequencies as in Fig.2.}
\label{FIG.4}
\end{figure}

When the initial temperature of the nonlinear medium rises, there are initially more uncorrelated thermal photons in the wave-packets. We expect intuitively that, for a given number of micro-modes and for a given interaction time, any entanglement should vanish at large enough temperatures. To exemplify this behavior we compute the degree of entanglement ${ \cal E_N}$, for the {it one-to-all} interaction in terms of the temperature $T$, assuming $\tau_i = 0.6978$ and for: $n=1$ panel (a), and $n=5$ panel (b). The same dimensionless frequencies as in Fig.2 are assumed.  We observe that entanglement decreases with growing temperature until it vanishes at some critical temperature $T_c= 4.38^{\circ} K $ for $n=1$ and $T_c=1200^{\circ} K $ for $n=5$. As mentioned above, the CM for the {\it pairwise} pattern coincides with the CM of the {\it one-to-all} pattern for $n=1$ and we conclude that, for a given interaction time, the critical temperature $T_c$ is independent of the number of modes in the {\it pairwise} case and increases with $n$ for the {\it one-to-all} case, contradicting our initial intuition that, more macroscopic are the wave-packets more classical should be their collective correlations.

\begin{figure}[htb]
\centering
\includegraphics{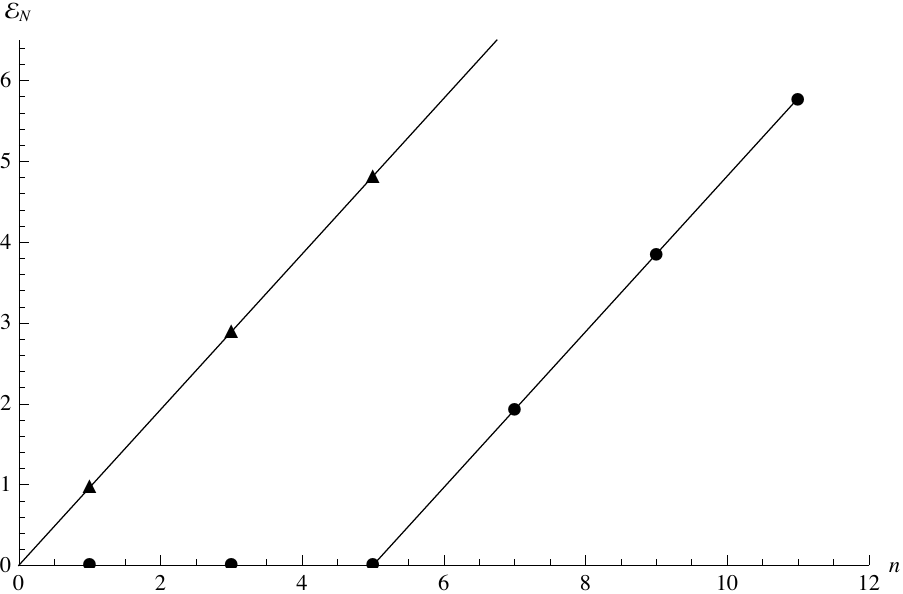}
\caption{\it 
Logarithmic negativity ${ \cal E_N}$ computed at the interaction dimensionless time $\tau_i =0.3324$ is displayed as a function of the number $n$ of micromodes for the initial vacuum state (triangles) and for a thermal state at equilibrium temperature $T= 30^{\circ} K$ (circles) and for the {\it one-to-all} interaction. The same dimensionless frequencies as in Fig.2 are assumed. The solid lines that fit the logarithmic negativity, are parallel straight lines whose equations are $ { \cal E_N}=0.963 \,\  n $ and $ { \cal E_N}=0.963\,\ (n-5)$.}
\label{FIG.5}
\end{figure}

To connect the physical {\it birth time of entanglement} $t_{\cal E}$ and the degree of entanglement ${ \cal E_N}$, with the coupling parameter $w$, we use the dimensionless time $ \tau = w t $. Let us assume that for a given temperature $T$ the dimensionless {\it birth time of entanglement} is $\tau_{\cal E}$, then when the coupling constant $w$ increases the real {\it birth time of entanglement} $t_{\cal E}$, decreases, meaning that stronger is the nonlinear interaction faster the {\it break-even} of energy is attained. In a similar way we show that the real interaction time $t_i$ needed to attain a given value of ${ \cal E_N}$, decreases with increasing $w$.

\begin{figure}[htb]
\centering
\includegraphics{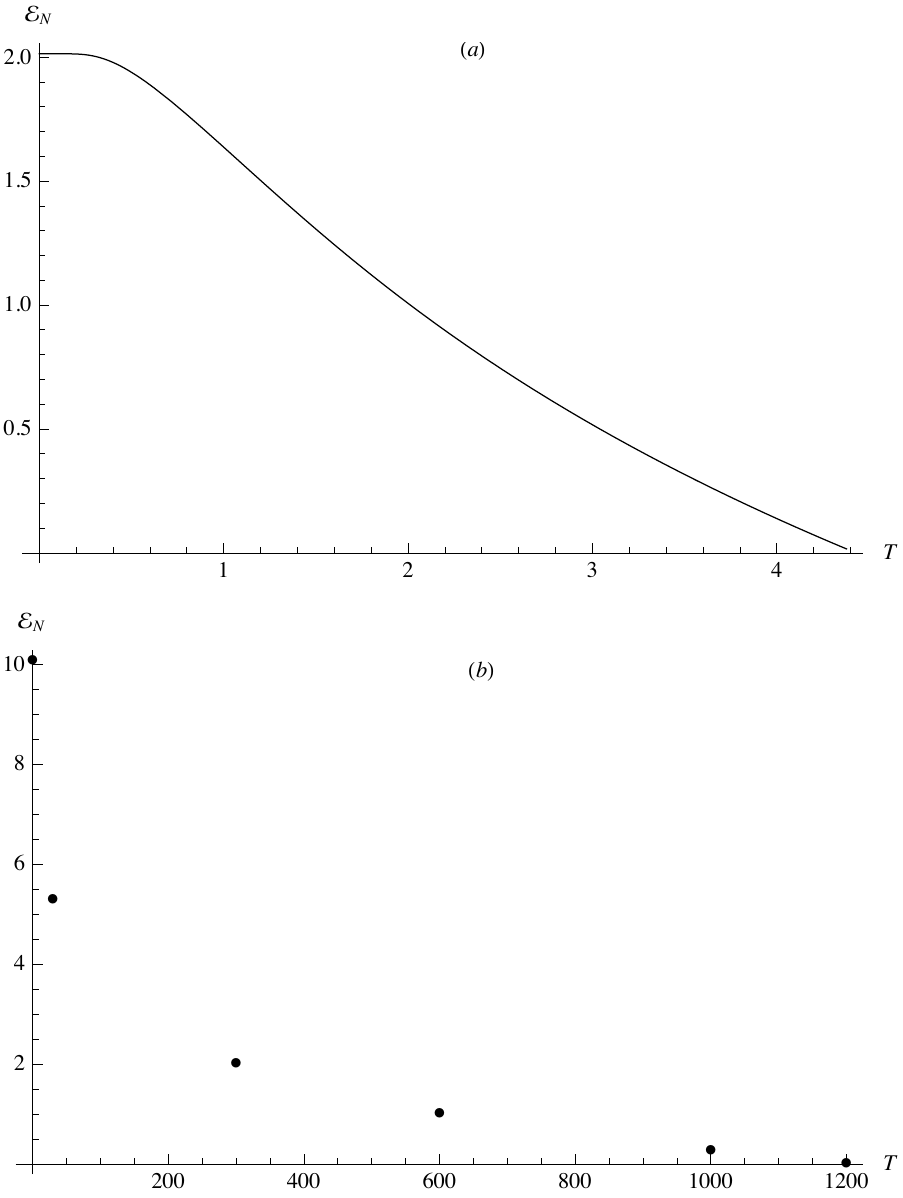}
\caption{\it 
Logarithmic negativity ${ \cal E_N}$, computed at the interaction dimensionless time $\tau_i =0.3324$, is displayed as a function of the temperature $T$. Panel (a)  $n=1$  and panel (b) $n=5$, for the {\it one-to-all} interaction. The same dimensionless frequencies as in Fig.2 are assumed.  The critical temperatures are $T_c=4.38^{\circ} K$ for $n=1$ and $T_c=1200^{\circ} K$ for $n=5$.}
\label{FIG.6}
\end{figure}

\subsection{Interpretation}

At a first glance the growth of the macroscopic degree of entanglement with $n$ and the decreasing of the BTE with $n$ are quite surprising behaviors. We would expect to see emerging a classical behavior with increasing number of constituents of the wave-packets,  this is, with their passage from microscopic to macroscopic systems. Instead, the collective quantum behavior is insensitive to the number of micro-modes in the {\it pairwise} interaction and is enhanced in the {\it one-to-all} interaction.

This behavior can be explained in the following way. When two micro-modes are added to each wave-packet (symmetric towards the central frequency ${ \bar \omega}_{\bf j}$) that contains initially $(n -2)$ modes, the interaction energy of the system increases roughly as $2 n \hbar w$ ($2n$ terms must be added to the interaction Hamiltonean) in the {\it one-to-all} interaction, and increases with $2 \hbar w$ in the {\it pairwise} interaction. However, the energy contained in the modes themselves (noninteracting Hamiltonean) increases with $2 \hbar ( {\bar \omega}_1 +{\bar \omega}_2)$ in both interaction patterns. The ratio between the interaction energy and, the non-interacting one, increases with $n$ in the pattern $G_1$ and is constant in the pattern $G_2$. An immediate result of this energy balance is that the {\it break-even} of energy (BTE) is attained earlier and the correlated pair production, that is responsible by the degree of entanglement, is enhanced by a factor of $n$ in $G_1$ and keeps constant in the pattern $G_2$.

This is expressed in a simple way by the degree of each vertex in the graphs $G_1$ and $G_2$. As a matter of fact, at a given temperature, the degree of entanglement for the {\it one-to-all} interaction is a linear function of the degree $n$ of any vertex of graph $G_1$. In the {\it pairwise} interaction the degree of any vertex of $G_2$ is $1$, independent of the number of vertices in the pattern, such as the degree of entanglement is insensitive to the number $n$ of micro-modes.

In conclusion, the number of constituents of the macroscopic systems is not {\it per se} the determinant factor when computing their degree of entanglement, the determinant ingredient is no doubt the type of interaction between their constituents.

\section{Summary and concluding remarks}

On the fundamental side our work demonstrates that, in principle, purely quantum correlations can be detected, at finite temperatures, by measuring a few number of appropriated collective observables of macroscopic systems.

To illustrate this behavior we studied the macroscopic quantum entanglement of the collective amplitudes of two wave-packets of finite spectral bandwidth, generated in a parametric down conversion process in two different approaches: the {\it pairwise} and the {\it one-to-all} interactions. 

We have shown that, in the {\it pairwise} pattern, the degree of collective entanglement does not depend on the number of micro-modes in the wave-packets and that, in the {\it one-to-all} pattern it increases linearly with the number of micro-modes. This quite surprising behavior not only contradicts our naive intuition that quantum correlations are averaged out in macroscopic systems as it emphasizes that, the complexity of physical interactions between the microscopic constituents, can enhance the degree of macroscopic bipartite entanglement. 

However, there is a minimum interaction time needed before the two wave-packets start to be entangled, the {\it birth time of entanglement}, which increases with increasing temperature. 

Moreover, for the same interaction time and the same number of micro-modes, the degree of entanglement decreases with increasing temperature, becoming zero at the critical temperature, where its first order derivative has a discontinuity. To associate this discontinuity with a possible quantum phase transition \cite{Vedral2007, Vedral2008(2)} is not clear within our model. It would be interesting, in a further work, to use the techniques developed in solid state physics to relate the polarization and the temperature of the nonlinear medium where the waves interact, with their degree of entanglement, in order to find signatures of quantum phase transitions.

Since this macroscopic entanglement is higher than the corresponding microscopic one and since the measuring apparatus only needs to detect the collective behavior of the wave-packets, it could be used as a resource to perform communication and computation tasks, at room temperature, in an efficient and technically more accessible way.

 \section{Appendix A: Microscopic evolution}
 
 In this Appendix we solve the Heisenberg equations of motion for the Bose operators of the micromodes of the wave-packets.
 
 The solution of the Heisenberg equations of motion for ${\hat a}_{1,k}$ and ${\hat a}_{2,-k}^{\dag} $ in the {\it pairwise} interaction, was derived in  \cite{Louisell1961}, and is given by
\begin{equation}\label{PDC3}
\hat{a}_{1,k} (t) = e^{-i \omega_{1,k} t} \left( \cosh wt \,\ \hat{a}_{10,k}  + \sinh wt \,\  \hat{a}_{20,-k}^{\dag} \right) 
\end{equation}
\begin{equation}\label{PDC4}
 \hat{a}_{2,-k}^{\dag} (t) = e^{i \omega_{2,-k}  t} \left(  \sinh wt \,\ \hat{a}_{10,k} + \cosh wt \,\  \hat{a}_{20,-k}^{\dag} \right)
\end{equation}
The zero stands for initial time.

For clarity of exposition let us rename the bosonic operators $\hat{a}_{1,k}$ in the signal wave-packet {\bf 1} and $\hat{a}_{2,l}  $ in the idler wave-packet {\bf 2} for the {\it one-to-all} interaction. The correspondence is the following: for modes of {\bf 1}: $\hat{a}_{1, -m} \rightarrow \hat{a}_{1} ;  \hat{a}_{1, -m+1} \rightarrow \hat{a}_{2} \,\ ;...; \,\  \hat{a}_{1, m} \rightarrow  \hat{a}_{2m+1}$ and for modes of {\bf 2}: $\hat{a}_{2,-m} \rightarrow \hat{b}_{2m+2}; \,\ ...\,\ \hat{a}_{2,m} \rightarrow \hat{b}_{4m+2}$.  
With this new notation the Hamiltonian ${\hat H }_1 $  becomes
\begin{equation}\label{2waves2}
{\hat H }_1 = \sum_{k=1}^{2m+1} \hbar \omega_{k} {\hat a}_{k}^{\dag} {\hat a}_{k} + \sum_{ l=2m+2}^{4m+2} \hbar \omega_{l} {\hat b}_{l}^{\dag} {\hat b}_{l} + i w \hbar \sum_{k=1}^{2m+1} \sum_{l=2m+2}^{4m+2} ( e^{- i( {\bar  \omega} _0 t + \varphi )}{\hat a}_{k}^{\dag}{\hat b}_{l}^{\dag} -h.c.) 
\end{equation}

The Heisenberg equations of motion for the vector operator ${\vec Y}= \\ 
\{ {\hat a}_1 ,  {\hat a}_2 ,..., {\hat a}_{2m+1}, {\hat b}^{\dag}_{2m+2 },  {\hat b}^{\dag}_{2m+3 },..., {\hat b}^{\dag}_{4m+2} \}^T$, under the Hamiltonian ${\hat H}_1$ are 
\begin{equation}\label{Heis}
\frac{d {\vec Y}(t)}{dt} = -i N(t) {\vec Y}(t)
\end{equation} 
where $N(t)$ is $2n \times 2n$ time dependent matrix. In order to obtain an autonomous system of equations we define new annhilation operators  ${\vec Z}= F(t) {\vec Y} $ where $ F(t)$ is a $2n \times 2n $ diagonal matrix whose elements are 

\[ F_{ij}(t)   =  \left\{
\begin{array}{cc}
e^{i {\bar \omega}_1 t}    & \mbox{ if $ i=j= 1,...,2m+1$ } \\
e^{-i {\bar \omega}_2 t} & \mbox{ if $ i=j= 2m+2,...,4m+2$ } \\
0 &  i \neq  j
\end{array}
\right. \]
The new bose operators obey the following autonomous system of linear differential equations
 \begin{equation}\label{Heis5}
\frac{d {\vec Z}(t)}{dt} =M {\vec Z}(t)
\end{equation} 
where
\begin{equation}\label{transition} 
M= \left(
\begin{array}{cc}
A & B   \\
B^{\dag} & A \\
\end{array}
 \right)
\end{equation}  
is the $2n \times 2n$ matrix.
 $A$ is a diagonal $n \times n$ matrix, whose elements are
\[ A_{ij}   =  \left\{
\begin{array}{cc}
i \delta [ m-( i-1)] & \mbox{ if $ i=j= 1,...,2m+1$ } \\
0 &  i \neq  j \\
\end{array}
\right. \] 
and $B$ is the $n \times n$ constant matrix whose elements are
\[ B_{ij}   = w ;   (i, j= 1,...,2m+1  ) \] 
The solution of (\ref{Heis5}) is
\begin{equation}\label{SHeis5}
{\vec Z}(t) =  S e^{Dt} S^{-1} {\vec Z}(0) = G(t) {\vec Z}(0) 
\end{equation} 
Where $D$ is the a diagonal matrix whose elements are the eigenvalues of the matrix $M$, and $S$ is the matrix whose columns are the eigenvectors of matrix $M$. 
Going back to the original bose operators ${\vec Y}(t) $, we obtain  
\begin{equation}\label{SHeis6}
{\vec Y}(t) = F^{-1}(t) G(t) {\vec Y}(0) = R(t) {\vec Y}(0) 
\end{equation} 
More explicitly, the time dependent operators ${\hat a}_k (t) $ and ${\hat b}^{\dag}_l (t) $ in the {\it one-to-all} interaction are given by,
\begin{equation}\label{motion00}
\hat{a}_{k} (t) = \sum_{j=1}^{2m+1} r_{kj}(t) \hat{a}_{j0}  + \sum_{j=2m+1}^{4m+2} r_{kj}(t) \hat{b}_{j0}^{\dag}
\end{equation}
\begin{equation}\label{motion01}
\hat{b}_{l}^{\dag}(t)  = \sum_{j=1}^{2m+1} r_{lj} (t)  \hat{a}_{j0}   + \sum_{j=2m+2}^{4m+2} r_{lj} (t) \hat{b}_{j0}^{\dag}
\end{equation}
 in terms of the elements $r_{ij}(t)$ of matrix $R(t)$. The subscript zero refers to initial time.
 
The canonical commutation relations obeyed by ${\vec Y}$ imply that,
\begin{equation}\label{bogoliubov3}
\sum_{j=1}^{2m+1}  |r_{kj} (t) |^2  - \sum_{j=2m+2}^{4m+2} |r_{kj} (t) |^2=1 \,\,\ ; \,\,\ \sum_{j=1}^{2m+1} |r_{lj} (t) |^2  - \sum_{j=2m+2}^{4m+2} |r_{lj} (t) |^2=-1
\end{equation}
$k=1,...,2m+1$ and $l=2m+2,...,4m+2$.

\section{Appendix B: Covariance matrix for the collective operators}

In this Appendix we compute the elements of the CM ${\bf \Sigma}$ of the collective operators. 

The collective time-dependent amplitudes in the {\it one-to-all} interaction are 
\begin{equation}\label{colt1}
{\hat A} (t) = \frac{1}{\sqrt {2m+1}}\sum_{k=1}^{2m+1} \hat{a}_{k} (t) =\sum_{j=1}^{2m+1} m_j (t) \hat{a}_{j0} + \sum_{j=2m+2}^{4m+2} n_j (t) \hat{b}_{j0}^{\dag} 
\end{equation}
\begin{equation}\label{colt2}
 {\hat B}^{\dag} (t) =\frac{1}{\sqrt {2m+1}} \sum_{l=2m+2}^{4m+2} \hat{b}_{l}^{\dag} (t) =\sum_{j=1}^{2m+1} t_j(t) \hat{a}_{j0} + \sum_{j=2m+2}^{4m+2} u_j (t) \hat{b}_{j0}^{\dagger}\end{equation}
 where the coefficients
\begin{equation}\label{Coefficient1}
 m_j(t) =\frac{1}{\sqrt {2m+1}}\sum_{k=1}^{2m+1}   r_{kj}(t) ;\,\,\    n_j(t) = \frac{1}{\sqrt {2m+1}}\sum_{k=1}^{2m+1}   r_{kj}(t)
  \end{equation} 
  \begin{equation}\label{Coefficient2}
 t_j(t) =\frac{1}{\sqrt {2m+1}}\sum_{l=2m+2}^{4m+2}   r_{lj} (t) ;\,\,\    u_j(t) = \frac{1}{\sqrt {2m+1}}\sum_{l=2m+2}^{4m+2}   r_{lj}(t)
  \end{equation} 
 are expressed in terms of the entries of the matrix $R(t)$.
The expressions for the elements of the symmetric collective covariant matrix for the initial thermal state, are easily computed 
\begin{equation}\label{sigma11}
\Sigma_{11} (t) = \Sigma_{22}  (t) =  \langle {\hat N}_1 (t) \rangle +\frac{1}{2}
\end{equation} 

\begin{equation}\label{N1}
\langle {\hat N}_1 (t) \rangle= \sum_{j=1}^{2m+1}   ({\bar n}_{j0}  +1)    |m_j (t)|^2   + \sum_{j=2m+2}^{4m+2} {\bar n}_{j0}   |n_j (t)|^2 
\end{equation}

\begin{equation}\label{sigma33}
\Sigma_{33}  (t) = \Sigma_{44}   (t) =  \langle {\hat N}_2  (t) \rangle +\frac{1}{2}
\end{equation} 

\begin{equation}\label{N2}
\langle {\hat N}_2 (t) \rangle= \sum_{j=1}^{2m+1}  {\bar n}_{j0}  |t_j (t)|^2   + \sum_{j=2m+2}^{4m+2}  ( {\bar n}_{j0}  +1)    |u_j (t)|2 
\end{equation} 

\begin{equation}\label{sigma12}
\Sigma_{12} (t) = \Sigma_{34}  (t) = 0
\end{equation}

\begin{equation}\label{sigma13}
\Sigma_{13} (t) =   \Re \left\{ \sum_{j=1}^{2m+1}  ( {\bar n}_{j0} + \frac{1}{2} )     m_j (t) t_j^{*} (t)+\sum_{j=2m+2}^{4m+2}  (  {\bar n}_{j0} + \frac{1}{2} )  n_j (t) u_j^{*} (t) )  \right\}
\end{equation} 

\begin{equation}\label{sigma24}
\Sigma_{24} (t) = - \Sigma_{13} (t) 
\end{equation} 

\begin{equation}\label{sigma14}
\Sigma_{14} (t) = \Im \left\{ \sum_{j=1}^{2m+1}  ({\bar n}_{j0} + \frac{1}{2} )     m_j (t) t_j^{*} (t)+\sum_{j=2m+2}^{4m+2}  (  {\bar n}_{j0} + \frac{1}{2} )  n_j (t) u_j^{*} (t) )  \right\}
\end{equation} 

\begin{equation}\label{sigma23}
\Sigma_{23} (t) =  \Sigma_{14} (t) 
\end{equation}

The coefficients of the CM for the {\it pairwise} interaction are
\begin{equation}\label{sigma11p}
\Sigma_{11} (t) = \Sigma_{22}  (t) =   \langle {\hat N}_1 (t) \rangle +\frac{1}{2}
\end{equation} 

\begin{equation}\label{N1p}
\langle {\hat N}_1 (t) \rangle=  {\bar N}_{10} \,\ {\cosh}^2 wt+  ( {\bar N}_{20}+1) \,\ {\sinh}^2 wt 
\end{equation}

\begin{equation}\label{sigma33p}
\Sigma_{33}  (t) = \Sigma_{44}   (t) =  \langle {\hat N}_2  (t) \rangle +\frac{1}{2}
\end{equation} 

\begin{equation}\label{N2p}
\langle {\hat N}_2 (t) \rangle= {\bar N}_{10} \,\ {\cosh}^2 wt + ( {\bar N}_{20} +1) \,\ {\sinh}^2 wt  
\end{equation} 

\begin{equation}\label{sigma12p}
\Sigma_{12} (t) = \Sigma_{34}  (t) = 0
\end{equation}

\begin{equation}\label{sigma13p}
\Sigma_{13} (t) =  \frac{1}{2}\sinh (2wt) \cos (2 {\bar \omega}_0 t) ({\bar N}_{10}+{\bar N}_{20} +1)
\end{equation} 

\begin{equation}\label{sigma24p}
\Sigma_{24} (t) = - \Sigma_{13} (t) 
\end{equation} 

\begin{equation}\label{sigma14p}
\Sigma_{14} (t) = \frac{1}{2}\sinh (2wt) \sin (2 {\bar \omega}_0 t) ({\bar N}_{10}+{\bar N}_{20} +1)
\end{equation} 

\begin{equation}\label{sigma23p}
\Sigma_{23} (t) =  \Sigma_{14} (t) 
\end{equation}

Where 
\begin{equation}\label{N10p}
  {\bar  N}_{10}= \langle {\hat N}_{1}(0)  \rangle  = \sum_{j=1}^{2m+1} \frac{{\bar n}_{j0}}{2m+1}     \,\,\  ,   \,\,\   {\bar  N}_{20}    =  \langle {\hat N}_2 (0)  \rangle=  \sum_{j=2m+2}^{4m+2}  \frac{{\bar n}_{j0}}{2m+1}  
\end{equation} 
are the initial mean number of photons in the wave-packets.

For both interaction patterns 
\begin{equation}\label{detgama}
\det \gamma (t) = - [ \Sigma_{13} ^2(t) +  \Sigma_{14}^2(t) ] \leq 0
\end{equation}

For the initial vacuum state, ${\bar n}_{j0} =0 $ for any $j$, therefore $\ {\bar N}_{10}= {\bar N}_{20} =0$ and $\det \gamma (t) = - {\sinh}^2 wt <0$ for $t>0$.

\end{document}